\documentclass[reprint,
amsmath,
amssymb,
floatfix,
superscriptaddress,
aps,
prx]{revtex4-2}
\usepackage{hyperref}
\usepackage{graphicx}
\usepackage{dcolumn}% Align table columns on decimal point
\usepackage{threeparttable}
\usepackage{bm}% bold math
\usepackage{color}
\usepackage{ulem}
\hypersetup{
    colorlinks=true,
    linkcolor=blue,
    citecolor=blue,   
   urlcolor=blue,
   }

\def\beqr{\begin{eqnarray}}
\def\eqnr{\end{eqnarray}}
\def\beq{\begin{equation}}
\def\bc{\begin{center}}
\def\ec{\end{center}}
\def\eqn{\end{equation}}
\def\bea{\begin{eqnarray}}
\def\eea{\end{eqnarray}}
\def\beq{\begin{equation}}
\def\eeq{\end{equation}}
\def\f{\frac}

\def\be{\beta}

\def\t{\tau}
\def\a{\alpha}
\def\on{k_{\rm on}}
\def\off{k_{\rm off}}

\def\r{\rho}

\def\s{\sigma}

\def\la{\langle}
\def\ra{\rangle}
\def\nn{\nonumber}

\def\d{\delta}
\def\p{\partial}
\def\l{\lambda}

\def\G{\Gamma}
\def\o{\Omega}

\def\g{\gamma}

\def\lb{\left(}
\def\rb{\right)}

\def\a{\alpha}
\def\d{\delta}
\def\p{\partial} 
\def\la{\langle}
\def\ra{\rangle}

\def\g{\gamma}

\def\hf{\frac{1}{2}}

\begin{document}
\title{UNC-104 transport properties are robust and independent of changes in its cargo binding}

\author{Amir Shee}%

\email[Amir Shee and Vidur Sabharwal contributed equally to this work]{}
\affiliation{Northwestern Institute on Complex Systems and ESAM, Northwestern University, Evanston, IL 60208, USA}
\affiliation{Institute of Physics, Sachivalaya Marg, Bhubaneswar-751005, Odisha, India}
\author{Vidur Sabharwal}
\email[Amir Shee and Vidur Sabharwal contributed equally to this work]{}
\affiliation{Department of Biological Sciences, Tata Institute of Fundamental Research, Mumbai, India}
\author{Sandhya P. Koushika}
\affiliation{Department of Biological Sciences, Tata Institute of Fundamental Research, Mumbai, India}
\author{Amitabha Nandi}%
\email[Author for correspondence:~]{amitabha@phy.iitb.ac.in}
\affiliation{Department of Physics, IIT Bombay, Powai, Mumbai 400076, India}
\author{Debasish Chaudhuri}
\email[Author for correspondence:~]{debc@iopb.res.in}
\affiliation{Institute of Physics, Sachivalaya Marg, Bhubaneswar-751005, Odisha, India}
\affiliation{Homi Bhabha National Institute, Anushakti Nagar, Mumbai 400094, India}
\date{\today}%
% PRX	≤  About 5% of article length & < 500 words
\begin{abstract}
Cargo distribution within eukaryotic cells relies on the active transport mechanisms driven by molecular motors. Despite their critical role, the intricate relationship between motor transport properties and cargo binding — and its impact on motor distribution — remains inadequately understood. Additionally, improper regulation of ubiquitination, a pivotal post-translational modification that affects protein degradation, activation, and localization, is associated with several neurodegenerative diseases.
Recent data showed that ubiquitination can alter motor-cargo binding of the Kinesin-3 motor UNC-104/KIF1A that transports synaptic vesicles.
To investigate how ubiquitin-like modifications affect motor protein function, particularly cargo binding, transport properties, and distribution, we utilize the PLM neuron of {\it C. elegans} as a model system. 
Using fluorescent microscopy, we assess the distribution of cargo-bound UNC-104 motors along the axon and probe their dynamics using FRAP experiments. We model cargo binding kinetics with a Master equation and motor density dynamics using a Fokker-Planck approach.
Our combined experimental and theoretical analysis reveals that ubiquitin-like knockdowns enhance UNC-104’s cooperative binding to its cargo. However, these modifications do not affect UNC-104's transport properties, such as processivity and diffusivity. Thus, while ubiquitin-like modifications significantly impact the cargo-binding of UNC-104, they do not alter its transport dynamics, keeping the homeostatic distribution of UNC-104 unchanged.

\end{abstract}

\maketitle

% PRX BODY	
\section{Introduction}
\label{sec_I}

Neurons are specialized cells that constitute the nervous system and are responsible for transmitting electrical and chemical signals across an organism. These cells typically have a cell body and neurites comprising a parallel arrangement of structurally polar microtubules with plus ends directed away from the cell body~\cite{ burton1981,heidemann1981} (see Fig.~\ref{schematic1}(a)). These microtubules are used by motor proteins (MPs) like kinesins~\cite{Vale1985} and dyneins~\cite{vallee1988} as tracks to transport cargo along axons. %~\cite{Hirokawa2010}. 
In axons, kinesins carry cargo anterogradely away from the cell body, whereas dyneins transport cargo retrogradely towards the cell body (e.g., Fig.~\ref{schematic1}(b)\,)~\cite{Vale2003, rai2013}.

The number of MPs on cargo affects its processive run length. Increasing the number of kinesins on cargo extends its duration of processive movement {\it in vitro}, but this is not always observed \textit{in vivo}~\cite{beeg2008,shubeita2008,wilson2021}.
{ 
Thus, the consequence of having more MPs on the cargo surface \textit{in vivo} remains unclear. In a neuron, optimal cargo transport away from the cell body may be aided by the processive motion of MPs. However, previous studies have shown that cargo transport is frequently bidirectional~\cite{martin1999}. Modeling reveals that bidirectional movement may aid in the circulation of cargo between the cell body and distal ends of the neuron~\cite{kuznetsov2022}. Such bidirectional movements can arise from either (i)~a tug-of-war between opposing MPs~\cite{soppina2009,hendricks2010}, or (ii)~coordination between opposing motors, where dynamic switching between opposing MPs can drive bidirectional motion~\cite{melanie2008,chowdary2015}. Indeed, both of these hypotheses have been validated in different contexts~\cite{munoz2022tug}. Thus, while processive MP movement is required for transporting cargo, optimal cargo transport may require precise control over MP binding to cargo and MP activation. 
{Indeed, MP-cargo binding and MP activation are under the control of several pathways that modify the MPs~\cite{guillaud2008, Mathieson2018,kevenaar2016}.}

{Optimal cargo distribution by MPs depends on their binding to cargo through the cargo-binding domain and their efficient transport, which relies on the processive motion of the motor domain along the filament~\cite{mitchell2012}.}
If MPs are not attached to their conjugate filaments, they can either diffuse freely or hitchhike after binding to a cargo~\cite{blasius2013}. Note that MPs must bind to cargo and attach to conjugate filaments simultaneously to mediate active cargo transport. Three different possibilities may arise~\cite{Vale2003, Klopfenstein2002, Rizalar2023, Kumar2010}: (i)~good cargo binding but weak filament binding, (ii)~good filament binding but weak cargo binding, and finally, (iii)~processive binding to both cargo and filament. All these possible scenarios are illustrated in Fig.~\ref{schematic1}(c) and may arise in different contexts. As an example, good cargo binding but weak filament binding is a typical feature of dynein, where an individual dynein molecule may not generate sufficient processivity and force for reliable cargo movement, but teams of dynein can work together to move a cargo~\cite{rai2013}. On the other hand, good filament binding but weak cargo binding has been observed in the slow transport of soluble cytoplasmic proteins -- they move by intermittent binding to motor proteins for directed motion, interspersed with dissociation and consequent diffusion~\cite{tang2013,roy2007}.
Finally, both synaptic vesicle proteins and autophagosomes bind reliably to motor proteins that move processively along filaments:  synaptic vesicle proteins move anterogradely, while autophagosomes move retrogradely~\cite{li1995,maday2012}.
Arguably, the last possibility,  where MPs processively bind to both cargo and filament, provides the most reliable transport toward the filament or axon termini but might reduce material availability closer to the cell body.  Note that the production of new MPs near the cell body and their subsequent degradation already ensures a possible steady state. However, the actual shape of the homeostatic distribution of MPs and their dynamics also depend on their filament processivity and related transport properties. 

Post-translational modifications (PTMs) of motor proteins (MPs) are essential for controlling their amount, activity, and cargo binding~\cite{guillaud2008,matthies1993, Gordon2001}. In particular, ubiquitination and ubiquitin-like PTMs play a key role in degrading and regulating MPs~\cite{Kumar2010, Sabharwal2024}. Disruptions in these PTMs are associated with various neurodegenerative diseases~\cite{perry1987,nakazawa2016,steffan2004}. Additionally, changes in molecular motors and cargo transport are implicated in the development and progression of these diseases \cite{gunawardena2003,lamonte2002,reid2002kinesin,zhao2001}. Therefore, understanding how ubiquitin and ubiquitin-like PTMs impact motor proteins can provide valuable insights into axonal cargo transport and highlight potential targets for therapeutic interventions in neurodegenerative diseases.

Recently, using a combination of live imaging of {\it C. elegans} neurons and theoretical analysis, we investigated how different ubiquitin-like modifications affect the cargo binding of Kinesin-3 MP UNC-104, a KIF1A ortholog~\cite{Sabharwal2024}. The results suggested that MPs may bind to cargo cooperatively, and the degree of cooperative binding depends on the ubiquitin-like modifications.

In this paper, we utilize RNAi-mediated knockdowns and live imaging of {\it C. elegans} to study the impact of different levels of ubiquitin-like modifications on the kinematics of the UNC-104 motor in the mechanosensory neurons, the posterior lateral microtubule (PLM) cells. %\alert
{With the help of a data collapse for cargo-bound MP distributions and theory, we characterize cooperative cargo binding and evaluate how the modifications influence this process.} We explore the impact on the homeostatic UNC-104 MP distribution along the axons and study the UNC-104 dynamics using localized FRAP. We analyze the experimental findings using a theoretical model. 
The experiments and theory allow us to uniquely determine UNC-104's effective diffusivity and drift.

The rest of the manuscript is organized as follows: Sec.\ref{worm_maintain} details the experimental system and methods; Sec.\ref{sec_cooperative_binding} covers the cargo-binding model and data-collapse of motor protein distributions; Sec.\ref{sec_transport} presents the theoretical model for motor protein dynamics along axons and uses experiments to evaluate transport properties; and Sec.\ref{sec_discuss} summarizes our findings and provides an outlook.

% === FIGURE 1 ================
\begin{figure*}[t!]
\includegraphics[width=16cm]{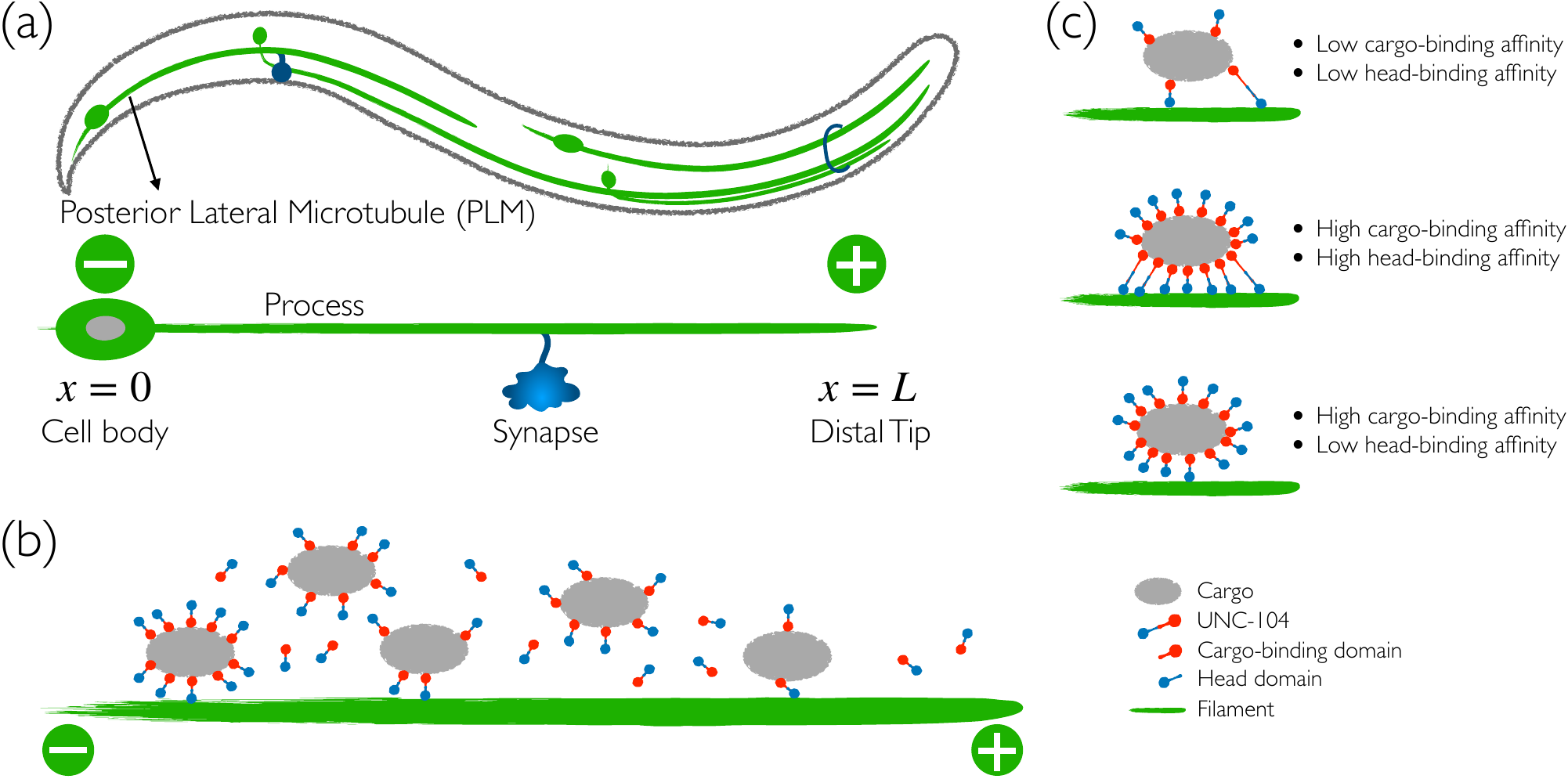}
\caption{Axonal transport in {\it C. elegans} (a) Schematic representation of the worm with the mechanosensory neurons indicated by green lines. The PLM neuron is further shown, indicating the cell body, axon, synapse, and neuronal process. The location of the cell body and the distal tip are labeled to be at $x=0$ and $x=L$, respectively. The plus and minus signs indicate the polarity of the aligned microtubules in the cell. (b)~Schematic showing multiple cargoes with attached UNC-104 MPs, represented by cargo-binding domain (in red) and head domain (in blue), which can bind to microtubules and walk along them in the attached state. When cross-linked to microtubules via MPs, cargo can be transported actively along the filament; otherwise, they diffuse detached from the filament. (c) Three possible scenarios of active cargo transport: (i) MPs with low cargo-binding affinity and low filament-processivity form weak cross-link between cargo and filament; (ii) MPs with high cargo-binding affinity and high filament-processivity lead to strong cross-link between cargo and filament, and (iii)~MPs with high cargo-binding affinity and low filament-processivity produce weak cargo-filament cross-link, represented by dense attachment of UNC-104 MPs to the cargo but not to the filament.
}
\label{schematic1}
\end{figure*}

\section{Material and Methods}
\label{worm_maintain}
\subsection{Worm maintenance}
%
%\noindent
\textit{C. elegans} was reared on NGM agar seeded with \textit{E. coli} OP50 using standard practices~\cite{Brenner1974}. For RNAi-mediated knockdown, NGM agar was prepared with 100 $\mu g\,  \mu L^{-1}$ ampicillin and 1 mM IPTG. These plates were seeded with the appropriate dsRNA [against either control empty vector (referred to as wild type), \textit{uba-1} or \textit{fbxb-65}] expressing \textit{E. coli} HT115 bacteria and incubated for 1 day at 20C before use. After 1 day, 4 \textit{C. elegans} young adults were placed and incubated at 20C. Their progeny at the L4 stage were then used for all experiments.

\subsection{Microscopy}
\label{microscopy}
%\noindent
To image UNC-104::GFP distribution, we used a Zeiss LSM 880 equipped with a 63x/1.4 N.A. oil objective at a frame size of 1024x1024 pixels leading to a pixel size of 88 nm using a high sensitivity GaAsP detector illuminated with a 488 nm argon LASER and imaged with a spectral filter set to a range of 493 to 555 nm. The entire neuron length was then imaged by tiling across six regions with a $\sim$10\% overlap at 5\% LASER power at 488 nm and 2x flyback line-scan averaging. Simultaneously, soluble mScarlet was imaged using a spectral filter from 585 to 688 nm with a 561 nm DPSS LASER at 5\% power at the same resolution. The images were automatically stitched using the Zen software.

\subsection{Steady state distribution of UNC-104 along the axonal length}
\label{steady_expt}
%\noindent
A line profile using a 3 pixels-wide spline fit polyline starting from the distal end of the PLM was traced up to the cell body, and the intensity and distance data were exported using FIJI~\cite{schindelin2012}. 
%
%\alert
{The exported data consists of the intensity profile along the entire axonal length for (a) UNC-104 tagged with GFP (UNC-104::GFP), and (b) mScarlet, which can diffuse freely and acts as a read-out of the axonal volume. Both (a) and (b) are obtained for the WT and ubiquitin-like modification knockdown animals. The exported data is further processed to obtain the final UNC-104 intensity profile, which is discussed in the next paragraph.}

{
In Fig.~\ref{fig2}, we show the experimental intensity profiles for the control (wild type) along the axonal length. The {normalized} average bare intensity profile $\la{I_{unc}}\ra$ of UNC-104::GFP is shown in Fig.~\ref{fig2}(a), and the corresponding {normalized} average mScarlet fluorophore profile $\la{I_{mscar}}\ra$ is shown in Fig.~\ref{fig2}(b). The intensity $I_{unc}$ is expected to be proportional to the number of UNC-104 per unit length of the axon. This can increase due to an increase in the density of UNC-104 or an increase in the local axonal volume at a fixed UNC-104 density. Note that the fluorophore profile gives a measure of the local volume of the cell along each axon. It varies as the width of the axon changes from the cell body to the terminal. In order to get the correct readout of the UNC-104 localization per unit volume, we take the ratio $I_{unc}/I_{mscar}$ of the intensity $I_{unc}$ of UNC-104::GFP to mScarlet intensity $I_{mscar}$, which provides insight into the distribution of UNC-104 MPs along the axonal length by mitigating variability due to axonal volume. The homeostatic profile of $\la{I_{unc}/I_{mscar}}\ra$ averaged over tens of cells is shown in Fig.~\ref{fig2}(c).
}

% === FIGURE 2 ===============
\begin{figure}[!b]
\includegraphics[width=8.6cm]{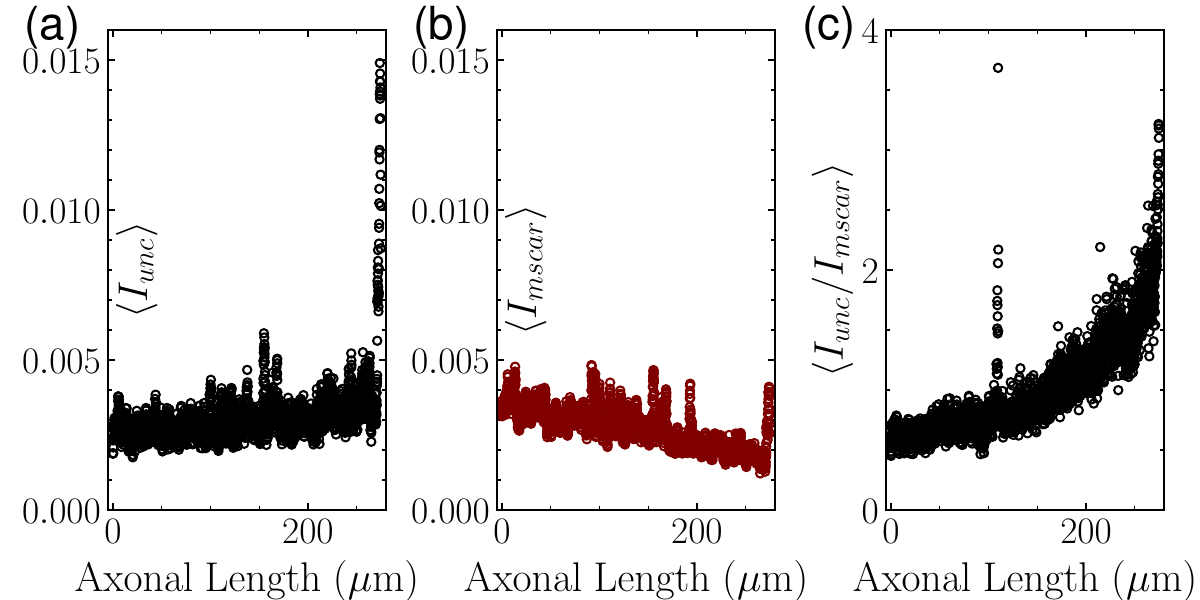}
\caption{(a) Variation of the average bare intensity profile of UNC-104::GFP $\la{I_{unc}}\ra$ and corresponding (b) fluorophore mScarlet intensity $\la{I_{mscar}}\ra$ for control (WT) experiments along the axonal length. (c) The relative intensity $\la{I_{unc}/I_{mscar}}\ra$ profile along the axonal length. All data are averaged over $n=15$ cells. 
}
\label{fig2}
\end{figure}

\subsection{Fluorescence Recovery After Photobleaching (FRAP) assay}
%\noindent
The imaging was done in an LSM 880 using a 40x/1.4 DIC M27 Oil objective at 1.5x zoom with a pixel size of 110 nm. Images were acquired every 500 ms in the ALM around 100 $\mu$m away from the cell body. The photobleaching was done after acquiring six pre-bleach images using a 488 nm diode laser at a region in the frame 50 $\mu$m in length using 80\% power (3 mW maximum power at objective) with five iterations. The imaging was done for at least 200 frames post-bleach.

%\noindent
Post-acquisition, movies were analyzed using FIJI~\cite{schindelin2012} by drawing a 3 pixels-wide spline fit polyline. The intensity profile at each time point was exported along with the ROI of bleaching and the bleach time point using a custom-built Python script. 
%\alert
{The analysis of the FRAP data using our model framework to obtain the diffusion coefficients is discussed in Sec.~\ref{sec:frap}}

\section{Theoretical Model to study the steady state distribution of cargo-bound kinesin motors}
\label{sec_cooperative_binding}
%\noindent

Recent theoretical analysis of cargo-bound UNC-104 motor proteins indicated a cooperative cargo binding mechanism~\cite{Sabharwal2024}.
We validate this cooperative binding model through a compelling data collapse by analyzing cargo puncta of different sizes, which correspond to varying numbers of bound motor proteins. Additionally, we assess how ubiquitin-like modifications influence the extent of this cooperative binding.

We consider the evolution for the probability $P(n,t)$ of a cargo bound to $n$ MPs at time $t$~\cite{Gardiner, Chaudhuri2011, Sabharwal2024}
\beqr
\label{meq}
\dot{P}(n,t)&=& u_{+}(n-1)P(n-1,t)+u_{-}(n+1)P(n+1,t) \nonumber \\
&&-[u_{+}(n)+u_{-}(n)]P(n,t).
\eqnr
Here, $u_{+}(n)$ and $u_{-}(n)$ denote the rates of MP binding and unbinding, cross-linking the cargo to the microtubule. We use a pairwise detailed balance at the steady state, $u_+(n-1)P_s(n-1)=u_-(n) P_s(n)$ and boundary condition $j_+= u_-(1) P_s(1)$ with $j_+$ a diffusion-limited rate of MP cross-linking. This leads to the exact steady state given by the recurrence relation
\bea
P_s(n)= \f{j_+}{u_-(n)} \prod_{m=1}^{(n-1)} \f{u_+(m)}{u_-(m)}
\eea
with the total number of motors $N=\sum_{n=1}^\infty n P_s(n)$. 

%\noindent
Cross-linked MPs can detach with a constant rate $\beta$ so that $u_-(n)=\beta n$. The presence of attached MPs can assist in further MP cross-linking such that $u_{+}(n)=a_{+}+b_{+}n$, within linear approximation. Here, $a_{+}$ denotes the basal attachment rate while $b_{+}$ quantifies the strength of cooperative binding. Using these expressions and expanding up to linear order in $1/N$ we obtain the following closed-form expression~\cite{Chaudhuri2011} 
\beq
\label{ss_formula}
H_s(n) = A\, n^\a e^{-\mu n}, 
\eqn
where $A=\f{j_+}{\be} \exp(\mu)$,  $\a=\f{a_+}{\be}-1$ and $\mu=1-\f{b_+}{\be}$.

The normalized distribution is $P_s(n)={\cal N}_n^{-1} H_s(n)$ with ${\cal N}_n = \int_0^\infty H_s(n) dn = A\, \mu^{-(\a+1)} \G(1+\a)$. This leads to
\bea
\tilde P_s(\mu n)=\f{\G(1+\a)}{\mu} P_s(n) = (\mu n)^\a \exp(-\mu n).
\label{norm_dist}
\eea

From the distribution of $P_s(n)$, we note that $\bar n = \a/\mu$ denotes both the mode and mean of the distribution, and its variance is given by $\s_n^2=2/\mu^2$.
We compare the theoretical distribution with the experimental results obtaining $\a \approx 1$ and $\mu_{WT}=7.33\times 10^{-3}$, $\mu_{uba1}=4.79\times 10^{-3}$ and $\mu_{fbxb65}=3.05\times 10^{-3}$~\cite{Sabharwal2024}.
Using these values, we plot the full distributions $\tilde P_s(\mu n)$ to obtain a data collapse in Fig.~\ref{fig3}. The plot of Eq.~(\ref{norm_dist}) using a black solid line agrees well with the experimental data as shown in Fig.~\ref{fig3}. The scaled quantity $a_+/\be\approx 2$ is independent of the RNAi and the cooperative binding strength $(b_+/\be)_{WT} < (b_+/\be)_{uba1} < (b_+/\be)_{fbxb65}$.

% === FIGURE 3 ===============
\begin{figure}[t!]
\includegraphics[width=8.5cm]{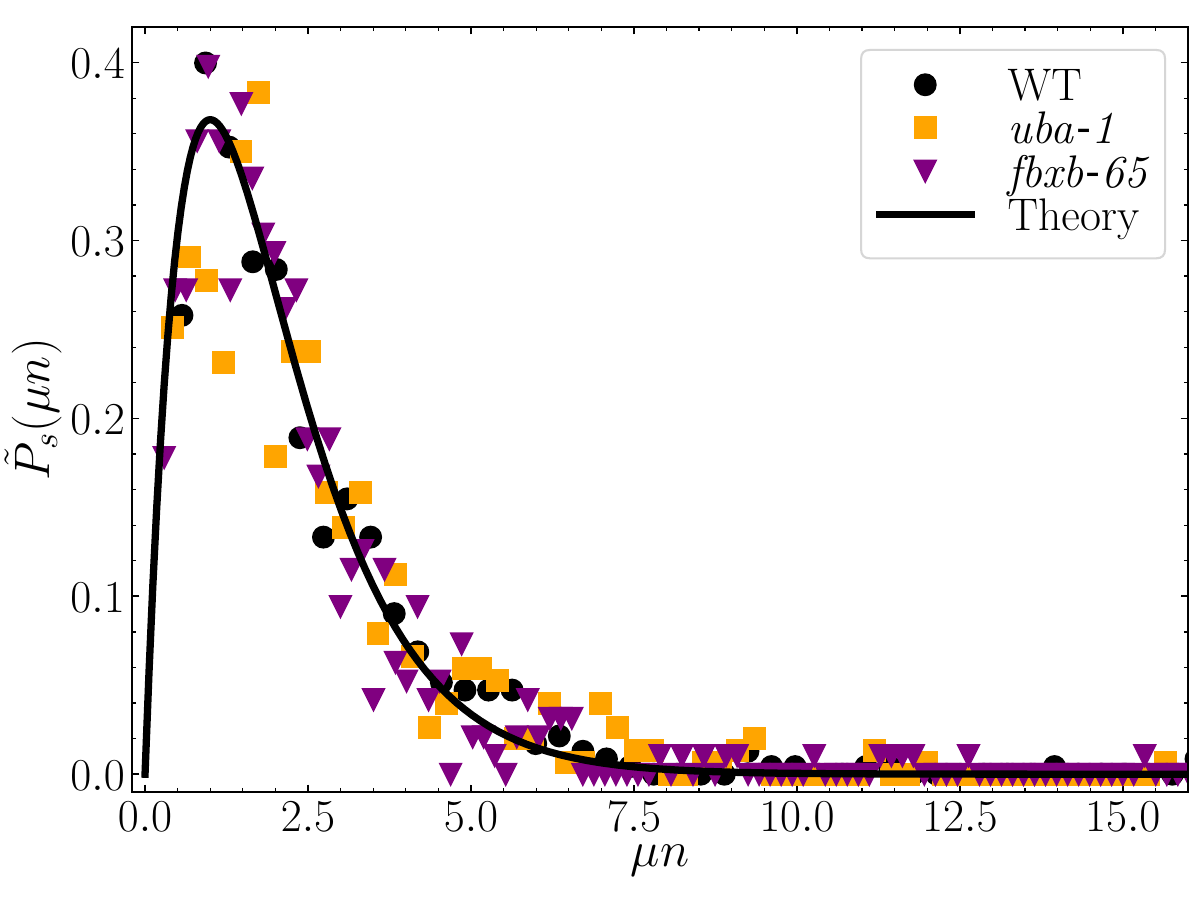}
\caption{Comparison of the steady-state distribution of the bound UNC-104 MPs between control (WT) experiments (black circles) and the \textit{uba-1} (orange squares) and \textit{fbxb-65} (purple triangles) knockdown cells. When scaled by corresponding $\mu$ values, the abscissa of the normalized intensity profiles shows a reasonable data collapse that agrees with the theory Eq.~(\ref{norm_dist}) with $\a=1$ plotted using the black solid line.
}
\label{fig3}
\end{figure}

Moreover, Eq.~(\ref{meq}), gives the evolution of average number of MPs $\la{n(t)}\ra$ cross-linking the cargo to microtubule, 
\beq
\label{avg_mp}
\dot{\la{n}\ra}=a_{+}+(b_{+}-\beta)\la{n(t)}\ra.
\eqn
Using the initial condition $\la{n(0)}\ra=n_0$, we obtain 
\beq
\label{avg_npsol}
\la{n(t)}\ra=\bar n - (\bar n-n_0)\exp(-t/\tau).
\eqn
As before, $\bar n=\a/\mu$ is the steady state average, and $\tau=1/\beta\mu$ is the relaxation time. This relaxation time is controlled by cooperative binding and should show a slower approach for {\it uba-1} and {\it fbxb-65} RNAi that have reduced UNC-104 ubiquitin-like modification.
We note that the typical relaxation time for the control (wild type) is $\tau_{WT}=1/\beta \mu_{WT} \approx 136.43~\mathrm{sec}$ considering the constant detachment rate of MPs $\beta = 1~\mathrm{sec}^{-1}$~\cite{Block2003}.
The relaxation time also determines the fluctuation correlation at the steady state,
\beq
\label{fluc_corr}
\la{\delta n(t)\delta n(0)}\ra = \s_n^2\exp(-t/\tau),
\eqn
where $\s_n^2=2/\mu^2$, the variance of steady-state  distribution $P_s(n)$. These predictions can be tested across future experiments. 

\section{Theoretical Model for evolution of UNC-104 density profile along axons}
\label{sec_transport}
The UNC-104 transport along microtubules aligned along axons can be approximated as a one-dimensional (1d) motion. 
The MPs can be either bound to microtubules and perform active, directed motion hydrolyzing ATP or diffuse in a passive manner when detached from the microtubule. Let us denote these bound and unbound fractions of MPs by $\r_b(x,t)$ and $\r_u(x,t)$, respectively. The bound fraction moves with an active drift velocity $v_0$. Here, we ignore the stochasticity of motion and the probabilities of back-stepping by ignoring the diffusion of the bound fraction. In contrast, the unbound MPs can only diffuse with diffusion constant $D_0$. Considering the binding and unbinding kinematics in terms of rates $\on$ and $\off$, we get
\bea
\p_t \r_b + v_0 \p_x \r_b &=& \on\r_u - \off\r_b, \\
\p_t \r_u - D_0 \p_x^2 \r_u &=& - \on\r_u + \off\r_b.
\label{den2}
\eea
At this stage, we ignored the synthesis and degradation of MPs, which will be incorporated later. 
The above equations can be rewritten in terms of the total MP density $\r=\r_b+\r_u$ and the density difference $m=\r_b-\r_u$. The evolution of $\r$ follows a conserved continuity equation. In contrast, the evolution of $m(x,t)$ gives 
$
\p_t m + v_0 \p_x \r_b + D_0 \p_x^2 \r_u = 2 (\on \r_u - \off \r_b).
$
In the presence of the source term on the right-hand side of the above equation, one can perform an adiabatic elimination in the hydrodynamic limit to obtain  
$\r_u/\r_b = \off/\on$.   
Using the processivity $\o = \on/(\on+\off)$ one can write $\r_b=\o \r$ and $\r_u=(1-\o)\r$. This leads to the conserved dynamics $\p_t \r + v \p_x \r - D \p_x^2 \r =0$ where the effective drift velocity and diffusivity of total MP density are $v = \o v_0$ and $D=(1-\o) D_0$. 

Now, we incorporate the other two source terms, the synthesis and degradation processes of MPs. 
We consider the synthesis at the cell body with rate $Q$ and a homogeneous degradation with rate $\g$. Thus, the  evolution of total concentration can be expressed as~(see Fig.~\ref{schematic1}),
\beqr
%\label{eq:dyn1}
\label{dyn1}
\p_t \r(x,t) =- v \p_x \r  + D \p_x^2 \r+Q \d(x) - \g \r.
\eqnr

The Dirac-delta function $\delta(x)$ ensures that the motor proteins are synthesized near the cell body ($x=0$). Remarkably, the above equation is an example of a stochastic resetting process that attracted tremendous recent interest in statistical physics~\cite{Evans2020}. However, unlike the typical stochastic resetting examples, in the present context, the number of particles is not exactly conserved. The mean value of the total number of MPs $n(t)=\int dx \rho(x,t)$ in the steady state $\bar n = Q/\gamma$ is determined by the synthesis and degradation rates $Q$ and $\g$.

\subsection{Steady state distribution}
%\noindent
We consider a reflective boundary condition at $x=L$ leading to $-D\p_x\r(x,t)|_{x=L}+v \r(x,t)|_{x=L}=0$. Thus, Eq.~(\ref{dyn1}) has the following steady-state solution~(see Appendix~\ref{app:1} for a detailed derivation):
\beq
\label{ss}
\r_s(x) =\f{Q \l_v\l e^{x/\lambda_v}}{2D\sinh(L/\l)}\Big{[}{\f{e^{-(L-x)/\l}}{(\l_v-\l)}+\f{e^{(L-x)/\l}}{(\l_v+\l)}}\Big{]}.
\eqn
Here two effective length scales $\lambda_v=2l_v$ and $\lambda=l_{\gamma}/[1+(l_{\gamma}/\l_v)^2]^{1/2}$ determine the steady-state profile. In these expressions, we used the characteristic length scales $l_v=D/v$ and $l_{\gamma}=\sqrt{D/\gamma}$.

A Laplace transform method can be employed to solve the dynamical equation Eq.~\eqref{dyn1} in the Laplace space. While an inverse transform to a closed-form expression could not be obtained, this solution allows us to determine the relaxation times towards the steady state. The slowest mode of time-scale $\g^{-1}$ is determined by the degradation rate $\g$ alone. The relaxation times towards steady-state for other relatively faster modes are $\g^{-1}\left[1+l_\g^2 \left( \f{1}{\l_v^2} + \f{\g \pi^2 n^2}{L^2} \right) \right]^{-1}$ for $n=1,2,3,\dots$; see Appendix~\ref{app:2}.

{The steady-state expression given by Eq.~\eqref{ss} may be used to fit the experimental profiles. However, we note that there are four independent parameters ($Q,D,v$, and $\g$) which makes this comparison difficult. We note that $\g\approx 10^{-4}$s$^{-1}$  as known from earlier studies for KIF1A \cite{Cohen2013, Fornasiero2018, Mathieson2018}. We further eliminate $Q$ using the total number of MPs
\beq
\label{normalization_ss}
\mathcal{N} = \int_{0}^{L}\r_s(x)dx=\frac{Q\lambda^2\lambda_v^2}{D(\lambda_v^2-\lambda^2)},
\eqn
to obtain a normalized steady-state distribution of MPs expressed in terms of the density profile $\r^{\mathcal{N}}_s(x) = \r_s(x)/{\cal N}$ and get
\beq
\label{norm_ss}
\r^{\mathcal{N}}_s(x) =\frac{(\lambda_v^2-\lambda^2)e^{x/\lambda_v}}{2\lambda\lambda_v\sinh{(L/\lambda)}}\Big{[}{\f{e^{-(L-x)/\l}}{(\l_v-\l)}+\f{e^{(L-x)/\l}}{(\l_v+\l)}}\Big{]}.
\eqn
Eq.~\eqref{norm_ss} has only two unknowns, namely $D$ and $v$. First, we estimate $D$ independently from the FRAP experiments, which are discussed in the next section. Second, we fit normalized experimental distribution using Eq.~\eqref{norm_ss} to get a single fit parameter $v$ with known values of $D$. Finally, we estimate values of $Q$ utilizing the fitting of non-normalized experimental distribution using Eq.~(\ref{ss}) with known values of $D$ and $v$.

\subsection{FRAP Analysis}
\label{sec:frap}
Although a numerical solution of Eq.~(\ref{dyn1}) for the evolution of the density profile can be obtained, the equation does not allow a simple closed-form solution. 
To analyze the experimental results, 
we use the following approach. In experiments, we normalize the evolution of the UNC-104 intensity profile by the homeostatic profile before photo-bleaching. The resultant profile is flat to begin with (before photo-bleaching). We observe the FRAP evolution with respect to the homeostatic profile. This evolution can be analyzed by simplifying the theoretical approach presented in Eq.~(\ref{dyn1}).

For this purpose, we consider the scaled evolution with respect to the theoretical steady-state, $\phi(x,t) = \r(x,t)/\r_s(x)$, over a small domain corresponding to the FRAP window. This follows,
\bea
\label{dyn2}
\p_t \phi(x,t) &=&  -v \p_x \phi(x,t)  + D \p_x^2 \phi(x,t)~,
\eea
locally, where we began by ignoring the synthesis and degradation terms for simplicity. This leads to a uniform steady state $\phi_s(x)=1$ in the finite domain. To model FRAP over a window of size $2a$, we analyze the evolution towards steady-state starting from  the initial condition
\beqr
\label{ic}
\phi(x_0,0) &=& 0~{\rm for}~ -a\leq x_0 \leq a , \nn\\
&=& 1 ~{\rm for}~ |x_0| > a.
\eqnr
 The Greens function corresponding to Eq.~\eqref{dyn2} is   
$G(x-x_0,t) = \f{1}{\sqrt{4\pi D t}} e^{-(x-x_0-v t)^2/4 D t}$. This, 
along with the initial condition in Eq.~(\ref{ic}),  leads to the solution  
\beqr
\phi(x,t) &=& \hf \left[2-{\rm erf}\left(\f{a+vt-x}{\sqrt{4Dt}}\right)-{\rm erf}\left(\f{a-vt+x}{\sqrt{4Dt}}\right)\right]~,\nonumber\\
\label{phi_t}
\eqnr
where ${\rm erf}(x)=\f{2}{\sqrt{\pi}}\int_0^x e^{-s^2}ds$. Now, bringing back the synthesis and degradation terms, the same analysis leads to a solution
\beqr
&&\phi(x,t) \nonumber \\
&&= \frac{e^{-\g t}}{2} \left[2-{\rm erf}\left(\f{a+vt-x}{\sqrt{4Dt}}\right)-{\rm erf}\left(\f{a-vt+x}{\sqrt{4Dt}}\right)\right]~\nonumber\\
&&+\frac{Q\Theta(t)}{\rho_0\g}(1-e^{-\g t}),
\label{phi_ft}
\eqnr
where $\r_0=\r_s(x=0)$. However, since the degradation rate $\g$ is known to be small, the intensity decay due to it over the short FRAP duration is negligible, and Eq.~(\ref{phi_ft}) reduces to Eq.~(\ref{phi_t}) apart from an additive constant $Q/\r_0\g$. The recovery of the profile $\phi(x,t)$ is mediated by effective diffusion $D$, degradation $\g$, and effective drift $v$.

% ==== BEGIN:FIGURE 4 ===============
\begin{figure}[t!]
\includegraphics[width=8.25cm]{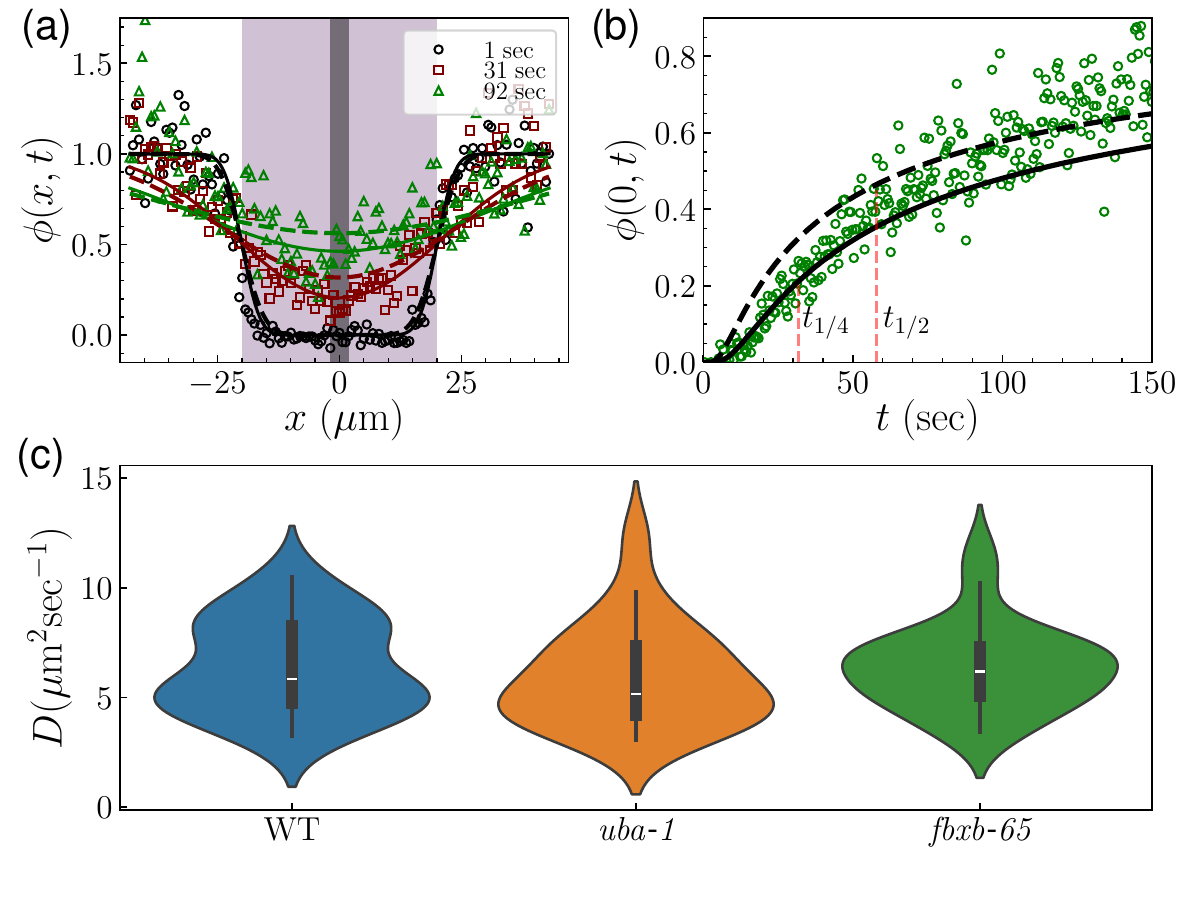}
\caption{Fluorescence Recovery After Photobleaching (FRAP):
(a)~The normalized intensity profile $\phi(x,t)$, scaled by the steady-state intensity, during FRAP at elapsed time $1$ sec (open circles), $31$ sec (open squares), and $92$ sec (open triangles). Symbols represent intensity profiles in the control (WT) cells.  The solid (using Eq.~(\ref{eq:diffusion_coeff_t1/4})) and dashed (using Eq.~(\ref{eq:diffusion_coeff_t1/2})) lines represent the plot of Eq.~(\ref{phi_t}) using the fitted diffusion coefficient $D$, the FRAP window size $w=2a$ and ignoring $v t \ll w$, to compare against experimental results. The region in violet shade indicates the range over which photobleaching is performed. A dynamical comparison between experiment, numerical solution, and theory is shown in the Supplemental Material Movie-1~\cite{Supply2024}.
(b)~Time evolution of the intensity $\phi(0,t)$ obtained by averaging over the narrow gray window of size $\epsilon$ indicated in (a). The diffusion coefficient $D$ can be calculated using half recovery time in Eq.~(\ref{eq:diffusion_coeff_t1/2}) and the quarter recovery time in Eq.~(\ref{eq:diffusion_coeff_t1/4}), both calculated from the experimental mean integrated intensity evolutions (open green points) similar to that shown in (b). The lines are the plot of Eq.~(\ref{eq:phi_time_evo}) using estimated $D$ by Eq.~(\ref{eq:diffusion_coeff_t1/2}) (dashed black line) and by Eq.~(\ref{eq:diffusion_coeff_t1/4}) (solid black line).
(c)~Violin plots of the diffusion constants obtained via $t_{1/4}$ using Eq.~(\ref{eq:diffusion_coeff_t1/4}), showing that the diffusivities do not depend on the ubiquitin-like knockdowns.
}
\label{fig4}
\end{figure}
% ==== END:FIGURE 4 ===============

%\as
{
We utilize the experimental intensity evolution during FRAP, after dividing by the homeostatic intensity profile, to estimate diffusion. We extract the diffusion coefficient by directly fitting Eq.~\eqref{phi_t} to the evolution of this intensity profile. Finally, by averaging over many realizations, i.e., different cells, we obtain the diffusion coefficient $D$ (Table~\ref{table1}).
}
While performing such fitting, one can neglect the small directed displacement $v t$ of the local density profile due to drift over the relatively short recovery time $t$ compared to the extent of the FRAP window $w=2a$, assuming $v t \ll w$. However, as it turns out, a strong intensity fluctuation makes it difficult to estimate $D$ reliably through this procedure. 

To reduce such systematic error, we use the evolution of an average intensity over a small window near the minimum intensity spot at the beginning of FRAP (Fig.~\ref{fig4}(a)). Setting $x=0$, $v t \ll a$, and $\g t\approx 0$, Eq.~(\ref{phi_t}) gives
\beq
\phi(0,t) = \left[ 1 - {\rm erf}\lb \f{a}{\sqrt{4Dt}}\rb \right].
\label{eq:phi_time_evo}
\eqn
We estimate the diffusion constant $D$ by computing the half recovery time $t_{1/2}$ of the center of the FRAP region defined over a small window of size $\epsilon=4\, \mu$m, where  $\epsilon\ll 2a$~{(considering $\epsilon$ window size as $10\%~\mathrm{of~FRAP~window~size}~2a$)} with $2a\approx 40 \,\mu$m. 
Thus, $\phi(0,t_{1/2})=(1/2) \phi(0,0)$ (see Fig.~\ref{fig4}(b)), we find the values of $D$ using 
\beq
D=0.275\f{(2a)^2}{t_{1/2}}.
\label{eq:diffusion_coeff_t1/2}
\eqn
{In our experiments, the recovery intensity turns out to be less reliable at longer times} due to large intensity fluctuations, and even determining $\t_{1/2}$ can be difficult for some experimental data. For them, we use $\phi(0,t_{1/4})=(1/4) \phi(0,0)$
(see Fig.~\ref{fig4}(b)) and the corresponding relation 
\beq
D=0.094 \f{(2a)^2}{t_{1/4}}.
\label{eq:diffusion_coeff_t1/4}
\eqn

% ==== BEGIN:FIGURE 5 ===============
\begin{figure*}[!t]
\includegraphics[width=17cm]{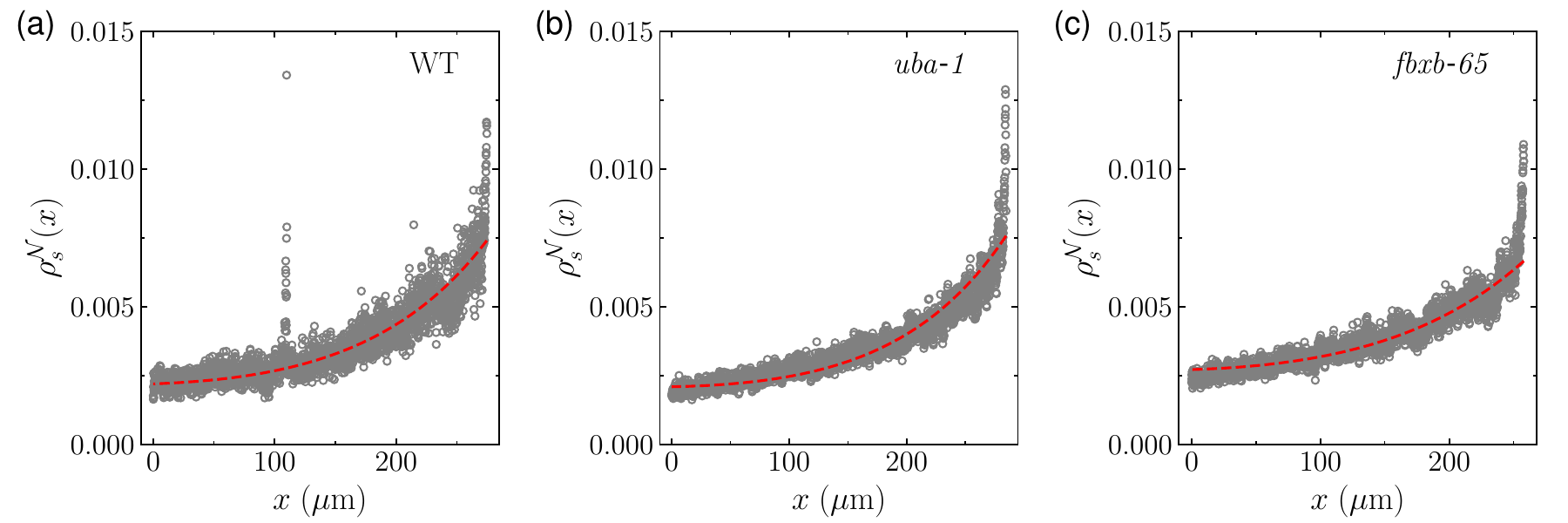}
\caption{Steady-state distribution of UNC-104 MPs: 
The normalized steady-state experimental distribution (open gray points) for control (WT) in (a), \textit{uba-1} knockdown cells in (b), and \textit{fbxb-65} knockdown cells in (c) are fitted with Eq.~(\ref{norm_ss}) (red dashed line) to estimate the single fitting parameter $\Omega$ as shown in Table-\ref{table1}. Here we used the known value of $\g$~($=10^{-4} \mathrm{sec}^{-1}$) and estimated diffusion coefficient $D$ from FRAP experiments (see Table-\ref{table1}).
}
\label{fig5}
\end{figure*}
% ==== END:FIGURE 5 ===============

We have estimated the effective diffusion coefficient $D$ using both $t_{1/2}$ and $t_{1/4}$ and found that diffusion coefficients were comparable (see Fig.~\ref{app:fig2}(c) and Table~\ref{table-sm:table1} in Appendix~\ref{app:2}). 
{Using the estimated $D$ values for the WT (both via $t_{1/2}$ and $t_{1/4}$), we compute $\phi(x,t)$ both numerically (solving Eq.~(\ref{dyn2})) as well as analytically (using Eq.~(\ref{phi_t})), neglecting the small advection. We show a comparison of the evolution at different time points of the FRAP experiments with these analytic estimates, which are shown in Fig.~\ref{fig4}(a) (see Supplemental Material Movie-1~\cite{Supply2024} for a comparison of the complete time evolution from experiment, numerical solutions and analytic expression). We also compare the theoretical time trajectories for $\phi(0,t)$ with the corresponding FRAP experiment; see Fig.~\ref{fig4}(b). We note that for this specific case, $\phi(0,t_{1/4})$ agrees better with the FRAP experiment at early times (up to $\lesssim$~60 sec) and deviates at large times when the recovery intensity is unreliable.  
%\alert
{For the \textit{fbxb-65} knockdown experiments, the $D$ values estimated using $t_{1/4}$ are statistically more reliable than the estimate using $t_{1/2}$, while for the WT and \textit{uba-1} knockdown, the statistics are equivalent (see Appendix.~\ref{app:3} for details).} The corresponding distributions of $D$ values, estimated using $t_{1/4}$, across different cells are compared between the WT and the two knockdown experiments in Fig.~\ref{fig4}(c) (see also the first row of Table \ref{table1}). Comparing the values of $D$ between the WT and the knockdown experiments, we notice that the relative variation in the value of $D$ across experiments is $\Delta D=(D_{WT}-D_{KD})/D_{WT}\leq 10\%$, where $WT$ and $KD$ stand for wild type and RNAi treated ubiquitin-like knockdowns respectively. 
}

%\noindent
The FRAP analysis thus suggests that cells with altered ubiquitin-like knockdowns of either \textit{uba-1} or \textit{fbxb-65} do not affect the MP's diffusional transport properties. Interestingly, due to the exponential form of the steady state profile $\r_s(x)$, a FRAP performed on top of this profile can lead to an apparent emergent retrograde bias in the recovery profile (see Appendix~\ref{app:4} and Movie-2 in~\cite{Supply2024}). This is a general physical effect that could be misleading and is, therefore, crucial to remember while analyzing any FRAP data. This is discussed in detail using our FRAP analysis in Appendix~\ref{app:4}.

\subsection{Active transport is crucial in determining the steady-state intensity profiles}
%\noindent
With the diffusion coefficients $D$ and the degradation rate $\g$ known, we can now determine the effective motor speed $v$ from the experimental steady-state distribution of the UNC-104 motors along the axon. To make a comparative study with our 1d theory, we must ensure that the steady-state distribution of MPs is defined purely as a one-dimensional density profile. 
This is discussed in Sec.~\ref{steady_expt}, where we corrected for the local volume variability of axons by scaling the experimentally obtained homeostatic intensity profiles for the MPs by the corresponding mScarlet fluorophore profiles for a given cell. This density further averaged over $n=15$ cells (see Fig.~\ref{fig2}(c)) gives a read-out for the steady-state local density profile $\r_s(x)$. Finally, we normalize the distribution $\r_s(x)$ to obtain $\r^{\mathcal{N}}_s(x)$ to eliminate the source term $Q$ and is used for comparison with the theoretical model.

Recall the effective motor speed is defined as $v=\o v_0$. The velocity $v_0$ of individual MPs obtained from live imaging of UNC-104::GFP-decorated puncta are estimated to be $\sim 1 \mu$m/s \cite{Sabharwal2024}. Assuming that the individual motor protein properties do not change, the unknown parameter $v$ can, therefore, be treated as a read-out for the value of processivity $\o$. Thus, to estimate $\o$ (via $v$), we further normalize the experimental steady-state distributions with respect to the spatially integrated intensity and compare them to Eq.~(\ref{norm_ss}). The results are shown in Fig.~\ref{fig5}(a), \ref{fig5}(b), and \ref{fig5}(c), which display a good fit in all the cases. Note that here we use only a single parameter fit to $v$ and hence obtain $\o$. The values of the processivity $\o$ obtained from these fits are reported in Table~\ref{table1}.
\begin{table}[b!]
\caption{Diffusion coefficient $D~(\mu \rm{m}^2 s^{-1})$ and processivity $\o$ with their relative errors estimated from the experiments.}
\label{table1}
\centering
\begin{ruledtabular}
\begin{tabular}{c c c c}
 & WT & \textit{uba-1} & \textit{fbxb-65}  \\
 \hline
$D$ & $6.41\pm 7 \%$ & $5.94\pm 8 \%$ & $6.37\pm 6 \%$ \\
$\Omega$ & $0.05\pm 3 \%$ & $0.05\pm 3 \%$ & $0.04\pm 2 \%$ \\
\end{tabular}
\end{ruledtabular}
\end{table}
%\noindent

{Several intriguing observations emerge from the fit results. First, it is evident that the transport parameters $D$ and $v$ remain largely unchanged by ubiquitin-like knockdowns, maintaining similar approximate values across all cases, including the wild type.}
Secondly, the processivity ($\o \approx 0.05$) is low, indicating that only $5\%$ of available motor proteins participate in transport. 
This also indicates that the intrinsic diffusivity $D_0=D/(1-\o)$ is mainly due to the detached fraction of MPs and is approximately equal to the measured values. 
%
%\blue
Additionally, we note that the integrated value of the MP density $\mathcal{N}=\int_0^L\r_s(x)$ (see Eq.~(\ref{normalization_ss})) remains almost unchanged. Since this term provides a read-out for the source term $Q$, we conclude that $Q$ is invariant to \textit{uba-1} and \textit{fbxb-65} knockdown cells~\footnote{$Q_{WT} \approx 0.035$, $Q_{uba-1} \approx 0.035$, $Q_{fbxb-65} \approx 0.033$.}.

% === FIGURE 5 ===============
\begin{figure}[t!]
\centering
\includegraphics[width=8.5cm]{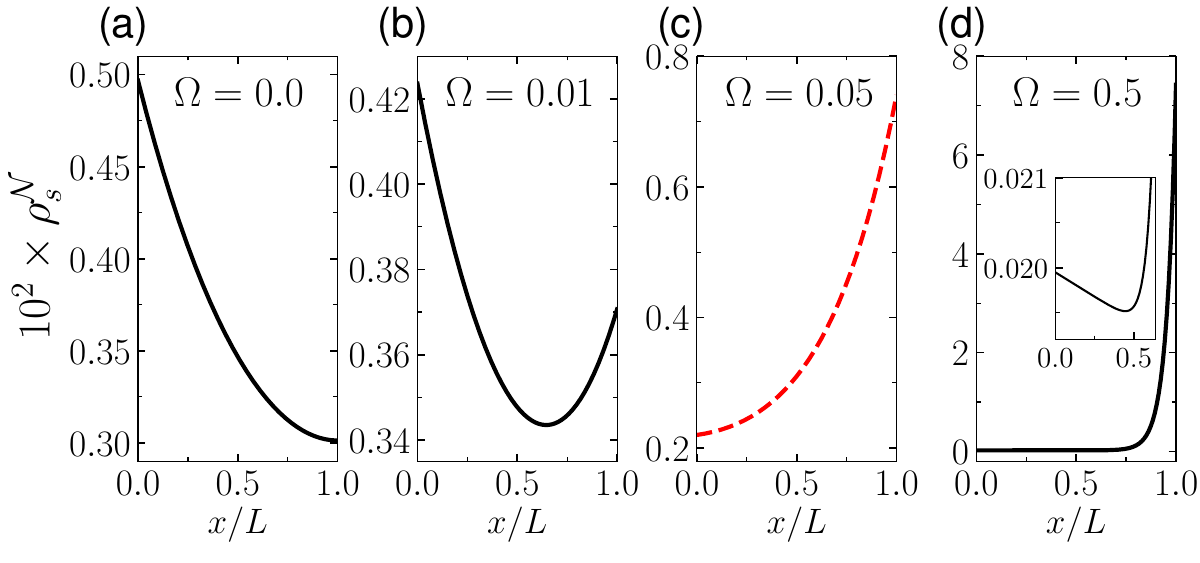}
\includegraphics[width=8.5cm]{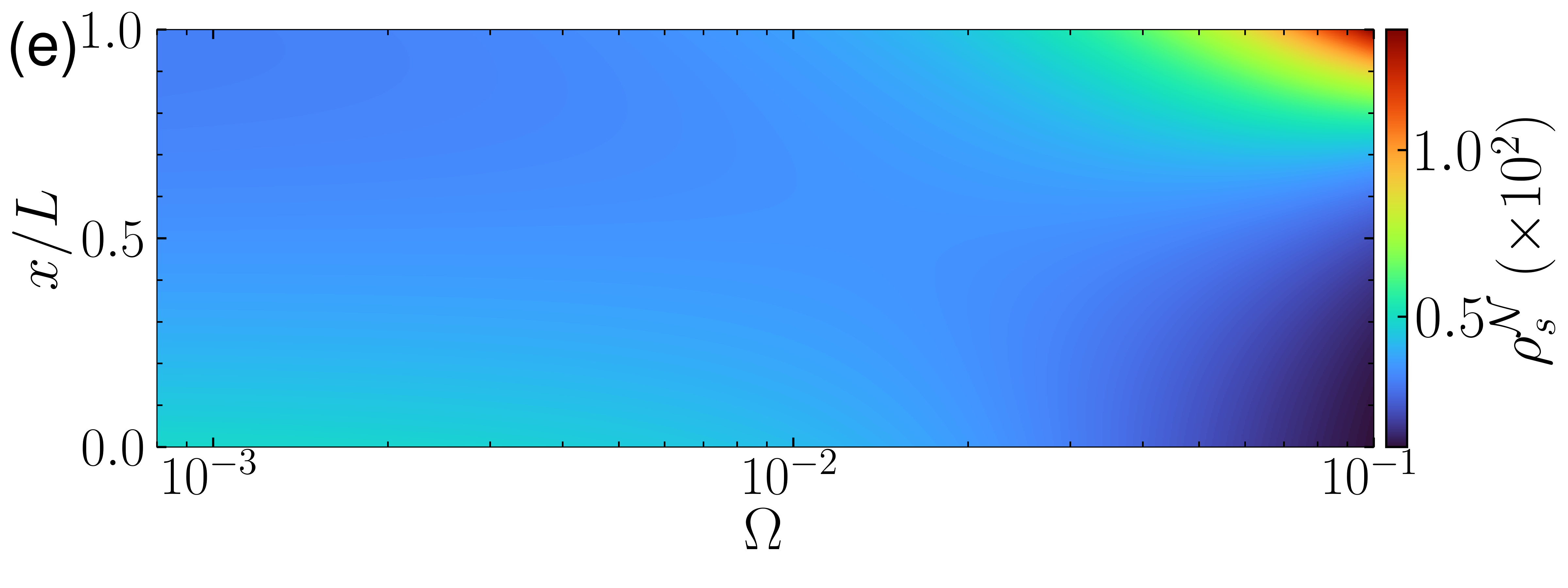}
\includegraphics[width=8.5cm]{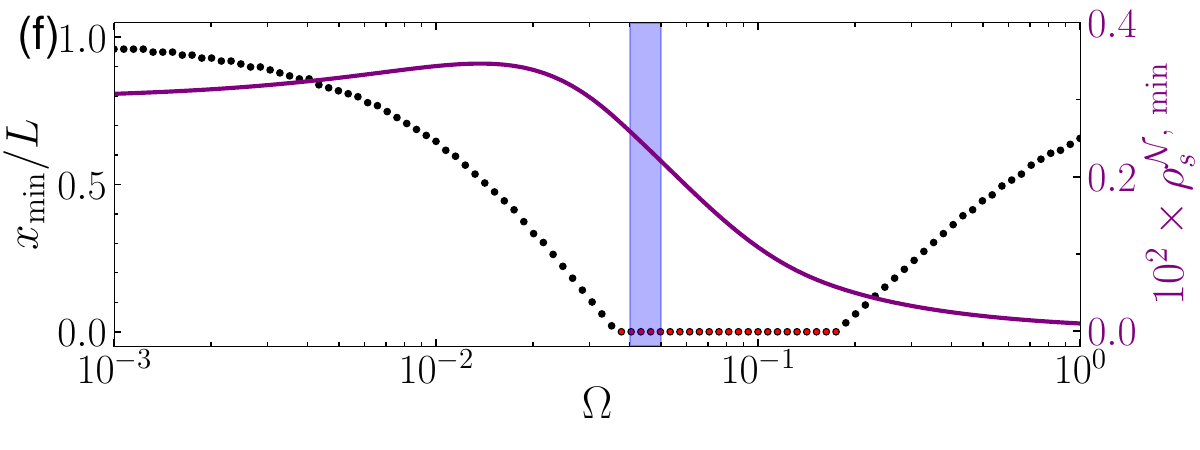}
\caption{
Normalized steady-state distribution of MPs {$\rho^{\mathcal{N}}_{s}(x)$ (see Eq.~(\ref{norm_ss}))} for different values of processivity $\Omega$ indicated in figures (a)-(d). The distribution in (c) corresponds to the WT (see Fig.~\ref{fig5}(a)). The zoomed-in plot in the inset in (d) highlights the minimum in the profile.
(e) Color map of  $\rho^{\mathcal{N}}_{s}$ as a function of position $x$ and processivity $\Omega$.
(f) The minimum of the MP distribution $\rho_{s}^{\mathcal{N}, \text{min}}$ (purple solid line) and its location on the axon at different processivities.  The red points mark the minima at zero, and the black points identify the minima at intermediate positions. The shaded region in (f) indicates the experimentally realized range $ 0.04 \leq \o \leq 0.05$. 
}
\label{fig6}
\end{figure}

The processivity $\o$ turns out to be the most important parameter regulating the steady-state distribution. From the expression in {Eq.~(\ref{norm_ss})}, we note that if $\o=0$, the length scale of the steady state exponential distribution is set by $l_{\g}$ and attains the maximum at the source (if $l_{\g}<L$) and the minimum at the posterior end (see Fig.~\ref{fig6}(a)). Upon slowly increasing $\o$, the minimum position of $\r_s(x)$ starts moving from $x=L$ to lower values of $x$ (see Fig.~\ref{fig6}(b)--(d)\,) until at $\o=0.05$, the {$\rho^{\mathcal{N}}_{s}(x)$} profile is already reversed and the length-scale is set by the competition between $l_{\g}$ and $l_v$ (see Fig.~\ref{fig6}(c)). 
%\dc{
This profile is similar to our experimental observation. It gets even sharper at higher $\o$, with most of the MPs accumulating near the distal end; see Fig.~\ref{fig6}(d). Moreover, as the inset of Fig.~\ref{fig6}(d) shows, 
the minimum of the profile shifts back to positive values at larger $\o$. The inversion of the density profile with the maximum shifting from proximal to distal end with increasing processivity is further illustrated using a heat-map in Fig.~\ref{fig6}(e). Fig.~\ref{fig6}(f) shows the non-monotonic variation of the minimum density and the location of this minimum with increasing $\o$.
%}
%

We can thus conclude that despite ubiquitin-like knockdowns of MP having a drastic effect on cooperative binding, the steady-state MP distribution regulated by the effective diffusivity and speed, which in turn are dependent on the effective processivity, remains completely unaffected.

\section{Discussion}
\label{sec_discuss}
Directed axonal transport of synaptic vesicles by MPs is crucial for the proper functioning of an organism, failure of which is associated with neurodegenerative disorders~\cite{gunawardena2003, lamonte2002, reid2002kinesin, zhao2001}. 
{ 
To ensure a robust and reliable transport of cargo, it is important to understand what controls the steady-state distributions of the MPs and what modifications can affect the motor's intrinsic transport properties}
As was shown earlier and further consolidated here using a convincing data collapse, knockdown of the E1 activating enzyme \textit{uba-1} or the E3 ligase \textit{fbxb-65} causes increased cooperative binding of UNC-104 to cargo. {In this work, we further show that despite such a change in cargo binding, the same ubiquitin-like knockdowns leave the UNC-104 distribution along the axon and their transport properties, like effective drift velocity and diffusivity, unchanged.} 

We quantified UNC-104 steady-state distribution in wild-type neurons and upon RNAi knockdowns of either \textit{uba-1} or \textit{fbxb-65}. The steady-state profile shows a larger accumulation of MPs in the distal end for all the cases. Normalization of the raw intensity data with the corresponding axonal volume showed a nice collapse of the data from different RNAi knockdowns on each other, indicating that the homeostatic distribution remains unchanged under ubiquitin-like knockdowns.

%\noindent
We proposed a theoretical model which clearly explains the nature of such a homeostatic profile. Synthesis, degradation, and transport, both directed drift and diffusion, control the steady-state density profile along the axon. 
%
%\noindent
Our model allowed us to estimate the drift and diffusion utilizing the homeostatic distribution of UNC-104 along the axons and analysts of FRAP results. 
We first estimated the diffusion coefficient $D$ from the FRAP experiments. Using the known $D$ values, we fitted the steady-state profiles to obtain estimates for the effective drift velocity $v$, which is a readout for the processivity $\o=v/v_0$. Our analysis clearly shows that both the diffusivity and the processivity remain unchanged under ubiquitin-like knockdowns.

{Since these kinesin-3 MPs can hitchhike on retrogradely moving cargo carried by dynein MPs, the kinetic properties measured using the overall fluorescent intensity from UNC-104 not only depend on their active motion along microtubules but also include the impact of such retrograde flux.}
We find that ubiquitin-like knockdowns increase cooperative cargo binding, thereby potentially increasing the impact of the hitchhiked retrograde motion. {Consistent with this, we previously observed a higher occurrence of UNC-104 retrograde movement under ubiquitin-like knockdowns~\cite{Sabharwal2024}.} Surprisingly, the effective kinetic properties of UNC-104 remain unchanged under the same ubiquitin-like knockdowns. Such independence suggests a subtle regulation nullifying the effect of potentially increased retrograde flux. Note that the cargo-bound MPs may support each other's microtubule processivity by sheer localization on the cargo.

\section*{Author Contributions} 
All authors contributed to the design of the research. VS performed experiments under the supervision of SPK. DC and AN designed the theoretical framework and analysis; AS performed numerical calculations; AS, AN, and DC analyzed the data. All authors discussed the results and wrote the manuscript.

\section*{Data availability }
All data that support the findings of this study are included within the article (and any supplementary files).

\section*{Acknowledgements}
AN and DC thank Madan Rao for insightful comments. DC thanks Sanjib Sabhapandit and Fernando Peruani for useful discussions. DC acknowledges research grants from the Department of Atomic Energy (OM no. 1603/2/2020/IoP/R\&D-II/15028) and Science and Engineering Research Board (SERB), India (MTR/2019/000750) and thanks the International Centre for Theoretical Sciences (ICTS-TIFR), Bangalore, for an Associateship. AN acknowledges SERB India (MTR/2023/000507) for financial support and thanks the Max-Planck Institute for the Physics of Complex Systems (MPIPKS), Dresden, for hospitality and support during summer visit in 2024. Research in SPK’s lab is supported by grants from DAE (OM no. 1303/2/2019/R\&D-II/DAE/2079; Project identification number RTI4003 dated 11.02.2020), PRISM (12-R\&D-IMS- 5.02-0202), and a Howard Hughes Medical Institute International Early Career Scientist Grant (55007425). AS acknowledges partial financial support from the John Templeton Foundation, Grant 62213.

\appendix

\section{Steady-state distributions}
\label{app:1}
We recast the governing equation~(\ref{dyn1}) in terms of current $J=-(D\p_x - v )\r$ and considering uniform degradation $\g(x) =\g$ constant,
\bea
\p_t \r &=& -\p_x J -\g \r~,
\label{eq:adv_diff_uniform_deg}
\eea
with the boundary condition of total current at $x=0$, $J_{x=0}=Q$ where $Q$ is the constant source and total current at $x=L$,  $J_{x=L}=0$.

The steady-state limit ($\p_t \r = 0$) gives
\bea
\p_{x}^2 \r_s - \f{1}{l_v} \p_x \r_s - \f{1}{l_\g^2} \r_s &=& 0
\label{eq:rho-s}
\eea
where the constants $l_v = D/v$ and $l_{\g}=\sqrt{D/\g}$. Substituting $\r_s \propto e^{\a x}$ in the Eq.~(\ref{eq:rho-s}) leads to the quadratic equation $\a^2-\a/l_v-1/l_\g^2=0$ with solution $\a = \left[l_\g \pm (4l_v^2 +l_\g^2)^{1/2}\right]/2l_v l_\g$. Thus, steady-state density,
\bea
\r_s (x) &=& e^{x/2l_v} \left[\mathcal{A} e^{\f{x\sqrt{4l_v^2 + l_\g^2}}{2l_v l_\g}}  + \mathcal{B} e^{\f{-x\sqrt{4l_v^2 + l_\g^2}}{2l_v l_\g}} \right]
\eea
We recast again with $\l_v = 2 l_v$ and 
$\l = 2l_v l_\g / \sqrt{4l_v^2 + l_\g^2}
=l_\g (1+l_\g^2/\l_v^2)^{-1/2}$. Thus,
\bea
\r_s (x) &=& e^{x/\l_v} \left[\mathcal{A} e^{x/\l}  + \mathcal{B} e^{-x/\l} \right]
\label{eq:rho-s1}
\eea
Now, the total current at $x=0$, $J_{x=0}=Q$ lead to
\bea
(\l -\l_v)\mathcal{A} + (\l +\l_v) \mathcal{B} &=& \f{Q\l \l_v }{D}~,
\label{eq:rho-s-x0}
\eea
and the total current at $x=L$, $J_{x=L}=0$ lead to
\bea
(\l -\l_v) e^{L/\l} \mathcal{A} + (\l +\l_v) e^{-L/\l} \mathcal{B} &=& 0~.
\label{eq:rho-s-xL}
\eea
Now, solving Eqs.~(\ref{eq:rho-s-x0}) and (\ref{eq:rho-s-xL}), we get
\bea
\mathcal{A} &=& \f{Q \l \l_v }{D (\l_v - \l  ) (e^{2L/\l} -1)}~, \nonumber\\
\mathcal{B} &=& \f{Q \l \l_v e^{2L/\l} }{D (\l + \l_v ) (e^{2L/\l} - 1)}~.\nonumber
\eea
Substituting $\mathcal{A}$ and $\mathcal{B}$ back in the Eq.~(\ref{eq:rho-s1}), we get
\bea
\r_s (x) = \f{Q  \l_v \l }{D (e^{2L/\l} -1)} e^{x/\l_v} \left[\f{e^{x/\l}}{(\l_v - \l)} + \f{e^{(2L-x)/\l}}{(\l_v + \l  )} \right]~,\nonumber\\
\label{eq:rho-ss}
\eea
which can be rewritten as:
\beq
\r_s(x) =\f{Q \l_v\l e^{x/\lambda_v}}{2D\sinh(L/\l)}\Big{[}{\f{e^{-(L-x)/\l}}{(\l_v-\l)}+\f{e^{(L-x)/\l}}{(\l_v+\l)}}\Big{]}.
\eqn
This is precisely Eq.~(\ref{ss}) of the main text.

\section{Laplace transform, singularities and slow dynamics}
\label{app:2}
The equation of motion for the density can be directly solved using Laplace transform $\tilde \r (x,s) = \int_0^t dt' e^{-s t'} \r(x,t')$ to get
%\begin{widetext}
\bea
\tilde \r (x,s) &=& \f{Q  \l_v \l(s)\, e^{x/\l_v} }{s D (e^{2L/\l(s)} -1)}  \left[\f{e^{x/\l(s)}}{(\l_v - \l(s))} + \f{e^{(2L-x)/\l(s)}}{(\l_v + \l(s)  )} \right] \nn\\
\label{eq:rho-s2}
\eea
%\end{widetext}
where $\l(s)=l_\g \left(1+\f{l_\g^2}{\l_v^2}+\f{s}{\g}\right)^{-1/2}$. It is easy to see that the above expression gives the steady-state result $\r_s(x)= \lim_{s\to 0} s \tilde \r (x,s)$.

The time dependence is given by the inverse Laplace transform, $\r(x,t)= \f{1}{2 \pi i} \int_{c-i\infty}^{c+i\infty} ds\, e^{s t}\, \tilde \r (x,s)$ with $c>0$ so that the function remains analytic on the right half of the complex plane. To perform the contour integration, we first need to analyze the structure of singularities of the integrand.  

Due to the presence of $\l(s)$ in the numerator, $\tilde \r (x,s)$ has a branch point at $s_b = - \g\left(1+\f{l_\g^2}{\l_v^2}\right)$. Moreover, it has simple poles at $(e^{2L/\l(s)} -1)=0$; using $1=e^{i\, 2n\pi}$ this gives simple poles at $s_n = s_b - \left(\f{\ell_\g}{L}\right)^2 \g n^2 \pi^2$ with $n=0,1,2,\dots$, all lying on the real axis and to the left of $s_b$.  
Further, as can be seen from Eq.~\eqref{eq:rho-s2}, $\tilde \r (x,s)$ has a simple pole at $s=0$, and at $s^* = s_b + \f{\ell_\g^2 \g}{\l_v^2} =-\g<0$ (using $\l_v = \pm \l(s)$) between $s=0$ and $s=s_b$.

Thus, the solution is
\beqr
\r(x,t)&=&\r_{s}(x) + \r_1(x)e^{s^* t} +  \sum_{n=0}^\infty \r_n(x) e^{s_n t}, \nonumber \\ %\f{1}{t^b} e^{s_b t}
&+& e^{s_b t} {\cal L}^{-1} \tilde \r(x, s+s_b)
\eqnr
so that $\r_1(x) = \lim_{s\to s^*} (s-s^*) \tilde \r(x,s)$, etc. 
The last term is the inverse Laplace transform of the frequency-shifted function, arising due to the branch point at $s_b<0$.
The slowest mode of evolution is $\lim_{t\to \infty}[\r(x,t) - \r_{s}(x)] \to \r_1(x)e^{-\g t}$ is controlled by the degradation rate $\g$. The reason for this becomes immediately clear by noting from Eq.~\eqref{dyn1} that the total quantity of MPs ${\cal N}=\int dx\, \r(x)$ evolves as $d {\cal N}/dt= Q-\g {\cal N}$.  

\section{Comparison of diffusivities from FRAP analysis}
\label{app:3}
%\noindent
In Fig.~\ref{app:fig1}, we show the scatter plot of diffusivities estimated using both $t_{1/2}$ (Eq.~(\ref{eq:diffusion_coeff_t1/2})) and $t_{1/4}$ (Eq.~(\ref{eq:diffusion_coeff_t1/4})) 
for control~(WT) along with \textit{uba-1} and \textit{fbxb-65} knockdown cells. 
% === APPENDIX FIGURE 1 ===========
\begin{figure}[t!]
\includegraphics[width=8.6cm]{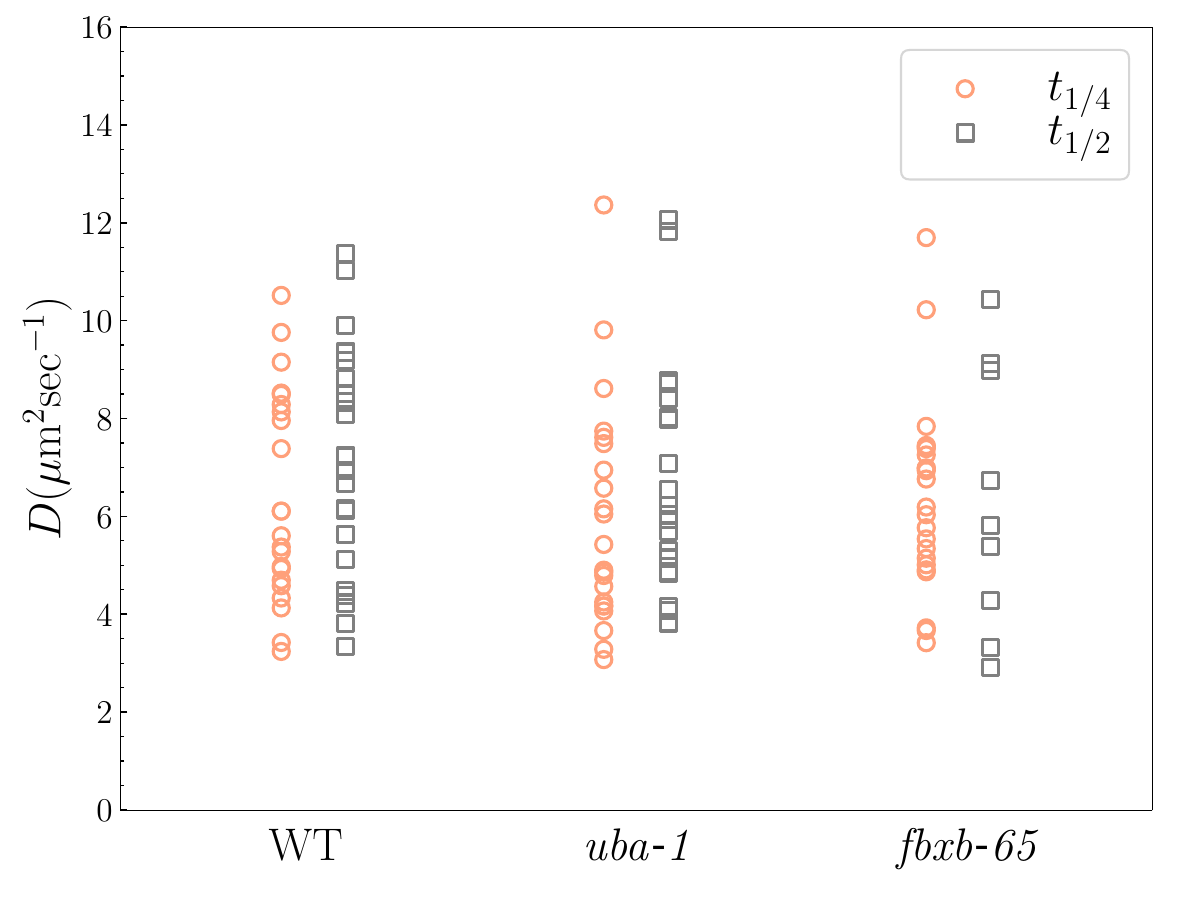}
\caption{Comparisons of the measured diffusion coefficients $D$ for control (WT) and the two types of RNAi cells using the quarter $t_{1/4}$ and half recovery times $t_{1/2}$. 
}
\label{app:fig1}
\end{figure}
The corresponding comparison for the mean values of effective diffusivities with their standard errors is shown in Table~\ref{table-sm:table1}. 
%\blue
{For both control~(WT) and \textit{uba-1} knockdown, $D$ is estimated by averaging over $22$ independent experimental realizations. However, for the case of \textit{fbxb-65} knockdown, $D$ measured via $t_{1/4}$ is averaged over all 23 realizations, whereas the one via $t_{1/2}$ was averaged over 9 realizations. This is because the $D$ estimate via $t_{1/2}$ was relatively unreliable in 14 cases due to large intensity fluctuations.}
\begin{table}[h!]
\caption{Comparison of diffusion coefficients $D$ (in $\mu$m$^2$ sec$^{-1}$) calculated using $t_{1/4}$ and $t_{1/2}$ for control~(WT) and cells with \textit{uba-1}, and \textit{fbxb-65} knockdowns. }
\label{table-sm:table1}
\begin{ruledtabular}
\begin{tabular}{ccc}
 RNAi & $D$~($t_{1/4}$) & $D$~($t_{1/2}$)\\ \hline
 WT & $6.41\pm 0.46$ & $7.02\pm 0.50$\\
\textit{uba-1} & $5.94\pm 0.49$ & $6.73\pm $ 0.50 \\
\textit{fbxb-65} & $6.37\pm 0.41$ & $6.33\pm 0.90$ \\
\end{tabular}
\end{ruledtabular}
\end{table}

\section{Further insights from the model: Apparent retrograde motion under FRAP}
\label{app:4}
%\noindent

%\alert
{To characterize the intensity recovery dynamics observed in the FRAP experiments, we numerically integrate Eq.~\eqref{dyn1} 
for an initial condition that is equivalent to the density profile just after FRAP. This is constructed as follows: starting with the steady-state profile defined in $[0,L]$, we choose a small window of size $w=2a=\Delta L$ around the center of the profile where the local density is set to zero~(Fig.~\ref{app:fig2}). The parameters are chosen corresponding to the WT case: we use the transport coefficients $D$ and $v$ from Table~\ref{table1} and the degradation rate $\gamma=10^{-4} \mathrm{sec}^{-1}$. We use the typical axon length $L=274.74~\mu\mathrm{m}$  and FRAP window size $w=2a=\Delta L=39.25~\mu\mathrm{m}$. To integrate Eq.~\eqref{dyn1}, we use time step $dt=0.0001~\mathrm{sec}$, set the source term $Q=0.035~\mathrm{sec}^{-1}$ at $x=0$ and use a reflecting boundary at $x=L$. The resultant MP distribution is finally normalized over the axonal length so that $\int_0^L \rho^{\cal N}(x) dx = 1$.}

%
% ==== BEGIN: APPENDIX FIGURE 2 ==============
\begin{figure}[!t]
\centering
\includegraphics[width=8.5cm]{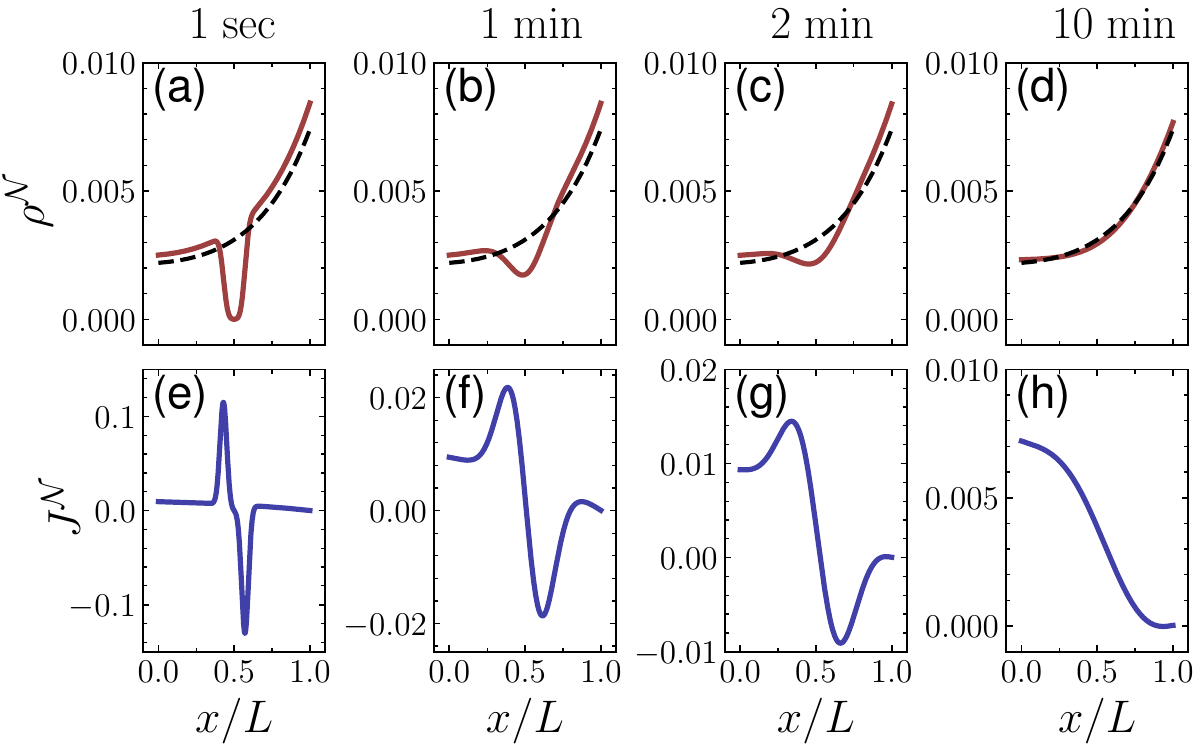}
\includegraphics[width=8.5cm]{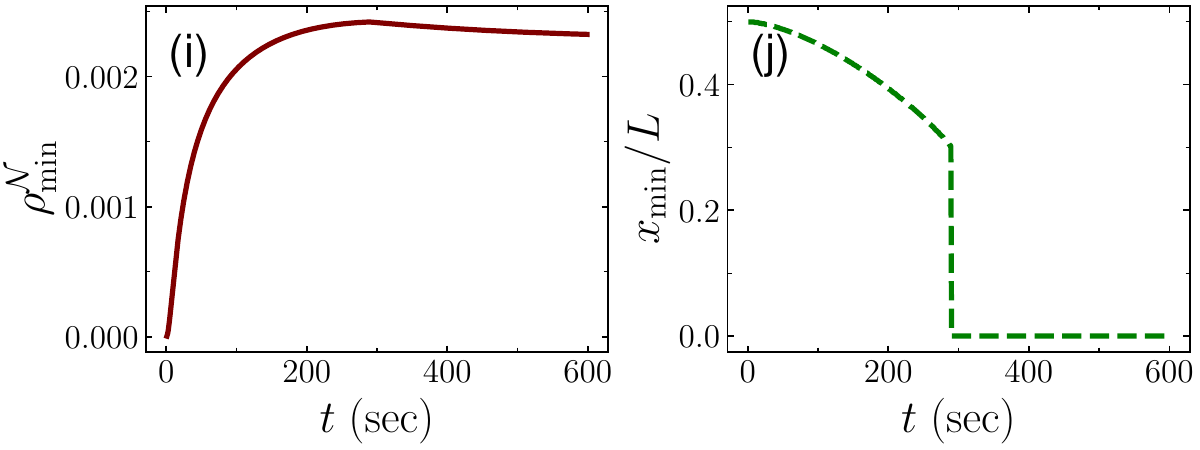}
\caption{
The spatiotemporal evolution for numerical calculation of FRAP of normalized spatial density $\rho^{\mathcal{N}}(x,t)$ (a-d) and normalized current $J^{\mathcal{N}}(x,t)$ (e-h) density profile of WT at times (a),(e) $t=1~$sec, (b),(f) $1~$min, (c),(g) $2~$min and (d),(h) $10~$min. The dashed black lines in (a)-(d) denote the normalized steady-state density profiles (Eq.~(\ref{norm_ss})) for comparison with the evolution. 
The plot of minimum density as a function of time is shown in (i), and the position of the minimum density is shown in (j).}
\label{app:fig2}
\end{figure}
% ==== END: APPENDIX FIGURE 2 ==============

~~

%\alert
{In Fig.~\ref{app:fig2}(a)-(d), the recovery of the density profile after FRAP is shown at four different times and the corresponding local currents $J^N(x,t)$ are shown in Fig.~\ref{app:fig2}(e)-(f) respectively. The approach toward the steady-state profile (shown by the dotted lines in Fig.~\ref{app:fig2}(a)-(d)) is evident from %Fig.~\ref{app:fig2}(a)-(d)
these figures, and at long times (see Fig.~\ref{app:fig2}(d)), the MP distribution almost overlaps with the steady-state profile. The recovery is further evident from the time evolution of local current $J^{\mathcal{N}}(x,t)$. Just after the FRAP, $J^{\mathcal{N}}(x,t)$ exhibit large positive and negative peaks corresponding to the two edges of the FRAP region (see Fig.~\ref{app:fig2}(e)). As time progresses, due to the recovery in depleted density, the peaks start receding until they disappear at long times (see Fig.~\ref{app:fig2}(f)-(h)) when $J^{\mathcal{N}}(x,t)$ approach to the steady-state value. It is important to note that at steady state, $J^{\mathcal{N}}_s(x)$ is not a constant due to the presence of a non-zero source $Q$ at one end. We also focus on the relaxation of the minimum density value $\rho^{\mathcal{N}}_{\mathrm{min}}$ (see Fig.~\ref{app:fig2}(i)) and its location $x_{\mathrm{min}}$ (see Fig.~\ref{app:fig2}(j)) along the axon. This indicates that the shape of the steady-state density profile almost recovers at around $t\sim 5~ \mathrm{min}$. Moreover, the decrease of $x_{\rm min}$ before vanishing captures an apparent retrograde motion of the location of FRAP minimum against the direction of MP flux. Similar behavior is observed in some of our FRAP experiments. This apparent retrograde movement is entirely due to the non-uniform shape of the steady-state distribution to which the perturbation under FRAP relaxes back. The relaxation of the profile during FRAP towards the approximate exponential steady-state profile led to the apparent retrograde motion of the FRAP center in the current example. {This is further demonstrated in Movie-2 presented in the Supplemental Material~\cite{Supply2024}, where we numerically show the FRAP region's evolution with the minimum density point labeled by a filled square.} Such apparent movement of the FRAP center is expected in any FRAP experiment on non-homogeneous steady-state profiles and, thus, must be taken into consideration while analyzing and interpreting such experimental results. }

%\newpage
%\UseRawInputEncoding
%\bibliography{reference}% Produces the bibliography via BibTeX.

\end{document}